\documentclass[aps,prb,twocolumn,superscriptaddress]{revtex4}
\usepackage[dvips]{graphicx}

\usepackage{bm}
\usepackage{amsmath}
\begin{document}
\title{
Charge dynamics of Ca$_{2-x}$Na$_{x}$CuO$_{2}$Cl$_{2}$\ as a correlated electron 
system with the ideal tetragonal lattice
}
\author{K.\ Waku}
\altaffiliation[Present address: ]{RISE, Waseda University, Tokyo 169-8555, Japan; PRESTO, 
Japan Science and Technology Corporation, Saitama 332-0012, Japan}
\affiliation{Department of Advanced Materials Science, University of Tokyo, Chiba 277-8562, 
Japan}
\author{T.\ Katsufuji}
\affiliation{Department of Physics, Waseda University, Tokyo 169-8555, Japan}
\affiliation{PRESTO, Japan Science and Technology Corporation, Saitama 332-0012, Japan} 
\author{Y.\ Kohsaka}
\affiliation{Department of Advanced Materials Science, University of Tokyo, Chiba 277-8562, 
Japan}
\author{T.\ Sasagawa}
\affiliation{Department of Advanced Materials Science, University of Tokyo, Chiba 277-8562, 
Japan}
\author{H.\ Takagi}
\affiliation{Department of Advanced Materials Science, University of Tokyo, Chiba 277-8562, 
Japan}
\affiliation{CREST, Japan Science and Technology Corporation, Saitama 332-0012, Japan}
\author{H.\ Kishida}
\affiliation{Department of Advanced Materials Science, University of Tokyo, Chiba 277-8562, 
Japan}
\affiliation{PRESTO, Japan Science and Technology Corporation, Saitama 332-0012, Japan}
\author{H.\ Okamoto}
\affiliation{Department of Advanced Materials Science, University of Tokyo, Chiba 277-8562, 
Japan}
\author{M.\ Azuma}
\affiliation{Institute of Chemical Research, Kyoto University, Kyoto 611-0011, Japan}
\affiliation{PRESTO, Japan Science and Technology Corporation, Saitama 332-0012, Japan}
\author{M.\ Takano}
\affiliation{Institute of Chemical Research, Kyoto University, Kyoto 611-0011, Japan}
\affiliation{CREST, Japan Science and Technology Corporation, Saitama 332-0012, Japan}
\date{\today}

\begin{abstract}
We report the reflectivity and the resistivity measurement of 
Ca$_{2-x}$Na$_x$CuO$_2$Cl$_2$ (CNCOC), which has a single-CuO$_2$-plane lattice with 
no orthorhombic distortion. The doping dependence of the in-plane optical conductivity spectra 
for CNCOC is qualitatively the same to those of other cuprates, but a slight difference between 
CNCOC and LSCO, i.e., the absence of the 1.5 eV peak in CNCOC, can be attributed to the 
smaller charge-stripe instability in CNCOC. The temperature dependence of the optical 
conductivity spectra of CNCOC has been analyzed both by the two-component model 
(Drude+Lorentzian) and by the one-component model (extended-Drude analysis). The latter 
analysis gives a universal trend of the scattering rate $\Gamma(\omega)$ with doping. It was 
also found that $\Gamma(\omega)$ shows a saturation behavior at high frequencies, whose 
origin is the same as that of resistivity saturation at high temperatures. 
\end{abstract}

%74.Superconductivity (for superconducting devices, see 85.25.-j)
%74.25.-q Properties of type I and type II superconductors
%*74.25.Fy Transport properties (electric and thermal conductivity, thermoelectric effects, etc.)
%*74.25.Gz Optical properties
%74.72.-h Cuprate superconductors (high-Tc and insulating parent compounds) 
%74.72.Bk Y-based cuprates
%74.72.Dn La-based cuprates
%74.72.Hs Bi-based cuprates
%*74.72.Jt Other cuprates, including Tl and Hg-based cuprates

\maketitle

\section{Introduction}
There has been a long history of discussions about the in-plane charge dynamics of cuprate 
superconductors. It is believed that the in-plane charge dynamics of cuprate superconductors is 
dominated by a small amount of holes introduced into a CuO$_{2}$ plane, which is 
theoretically represented by a tetragonal lattice of Cu$^{2+}$ ions ($3d^{9}$) with strong 
on-site coulomb repulsion. However, most of the cuprate superconductors have other 
characteristics that make the system away from such a simple two dimensional tetragonal lattice. 
First, there is often a different type of orthorhombic distortion in each system: 
La$_{2-x}$Sr$_{x}$CuO$_{4}$ (LSCO) has a buckling of 
CuO$_{6}$ octahedra,\cite{Fleming87} YBa$_{2}$Cu$_{3}$O$_{7}$ (YBCO) has CuO 
chains between CuO$_{2}$ planes, and Bi$_{2}$Sr$_{2}$CaCu$_{2}$O$_{8}$ (BSCCO) 
has the anisotropic modulation of BiO layers,\cite{Shaw88} all of which introduces 
orthorhombicity into the systems. Second, there is an instability of the stripe formation in the 
CuO$_{2}$ plane, which is particularly strong in LSCO.\cite{Tranquada95} Although it is not 
established whether such a stripe instability is an intrinsic nature of the tetragonal lattice, it is 
experimentally shown that the orthorhombic distortion largely affects the stripe formation in 
LSCO.\cite{Yamada99} This instability also complicates the system and its physics. Finally, 
YBCO and BSCCO has a bilayer structure of CuO$_{2}$ planes, and it is known that the 
inter-bilayer coupling cannot be ignored in such systems.\cite{Shen01} This difference between 
single-layer LSCO and bilayer YBCO or BSCCO makes it difficult to compare their charge 
dynamics in a quantitative way. 

Reflectivity measurement is a powerful technique to investigate the charge dynamics of metals 
and has been used for the study of both the in-plane and the out-of-plane charge dynamics in 
cuprate superconductors. As an overall feature, the doping dependence and the temperature 
dependence of the in-plane optical spectra are similar in all systems, YBCO,\cite{Cooper93} 
BSCCO,\cite{Terasaki90} and LSCO.\cite{Uchida91} Namely, upon doping, the peak around 
2 eV in the optical conductivity spectrum, which corresponds to the charge-transfer (CT) 
excitation between the Cu $3d$ and the oxygen $2p$ levels, decreases in its intensity whereas a 
quasi-Drude peak, which arises from the itinerant motion of the carriers, evolves below 1.0 eV. 
However, several details are different between the spectra of these systems. It was pointed out 
that the shape of the quasi-Drude peak below 1.0 eV is slightly different between these three 
systems: The quasi-Drude spectrum of LSCO has a dip around 0.1 eV and can be separated into 
a sharp Drude component below 0.1 eV and a Lorentzian above it, whereas that of BSCCO and 
YBCO is more smooth and does not look like the sum of two components.\cite{Tajima99} It 
was also pointed out that there is a peak existing between the CT excitation and the quasi-Drude 
peak in the LSCO spectra around 1.5 eV,\cite{Uchida91} which is absent in other two systems. 
These differences should come from the difference in the crystal structure as described above, 
but it has yet to be understood how the deviation from a tetragonal lattice affects the in-plane 
charge dynamics. 

Ca$_{2-x}$Na$_{x}$CuO$_{2}$Cl$_{2}$ (CNCOC) is one of the best systems in that sense 
to investigate the charge dynamics of the correlated electron system with a purely tetragonal 
lattice. This compound has a single-CuO$_{2}$-plane structure with apical chlorine 
ions\cite{Hiroi96} instead of apical oxygen ions in LSCO. Since the (Ca,Na)Cl plane 
separating two CuO$_{2}$ planes has a more ionic character than the (La,Sr)O plane in LSCO, 
it is expected that the coupling between two adjacent CuO$_{2}$ planes is much smaller in 
CNCOC than in LSCO. In addition, unlike LSCO, there is no buckling distortion of the 
octahedral network in CNCOC, thus being a simple tetragonal structure.\cite{Hiroi96} These 
two characteristics make CNCOC the best system representing the electron correlation in the 
purely tetragonal lattice. Previously, it was difficult to make single crystals of CNCOC because 
of the necessity of using high pressure even for making polycrystalline samples. However, 
recent progress in making single crystals under high pressure has overcome this 
obstacle,\cite{Kohsaka02} and now a series of single crystals with various doping level in 
CNCOC can be grown, which is large enough in size for resistivity and reflectivity 
measurement. In this paper, we report the resistivity and reflectivity measurement of CNCOC. 
In particular, we focus on the doping and temperature dependence of the in-plane charge 
dynamics in the normal phase studied by optical measurement. 

\section{Experimental}
Single crystals of CNCOC were grown by a flux method under high pressures. The details of 
crystal growth have already been published in Ref. \onlinecite{Kohsaka02}. Since the Na 
doped samples ($x > 0$) are highly hygroscopic, a special attention was paid not to expose the 
sample to the air in the preparation and measurement. The in-plane resistivity was measured by 
a standard four-probe technique, while the out-of-plane resistivity was measured by a 
quasi-Montgomery technique. In both cases, evaporated gold was used as the electrodes. The 
measurements were performed in the vacuum condition with a sample holder that was specially 
designed not to expose the sample to the air during the preparation and measurement. The 
reflectivity spectra were measured on the cleaved surface, which was prepared in the 
argon-filled glove box. We used a Fourier-type interferometer between 70 meV and 1.2 eV and 
a grating type spectrometer between 0.75 eV and 5 eV. The size of the sample we measured was 
1mm $\times$ 1mm at best, and the optically flat area is much smaller than that. Thus, all the 
measurements were done under the microscope attached to the spectrometer, with a typical spot 
size of 80 $\mu$m $\times$ 80 $\mu$m. Because of this size limitation, the measurement in 
the far-infrared region (below 70meV) cannot be made with our measurement system. For the 
measurement at room temperature, the sample was placed in a sealed small box filled with 
argon gas equipped with an optical window. Al mirror was also placed adjacent to the sample as 
a reference. We measured the reflectivity at low temperatures between 70 meV and 1.2 eV with 
a conduction-type cryostat in a vacuum condition. To obtain the absolute value of the reflectivity, 
we used the spectrum at room temperature, which was separately measured as described above, 
as a reference. We also measured the reflectivity of the undoped sample in the energy range 5 - 
34 eV using the synchrotron source at the Institute for Molecular Science (UV-SOR). Optical 
conductivity spectrum was calculated from the measured reflectivity spectrum using 
Kramers-Kronig relation. We used Hagen-Rubens extrapolation for $\hbar\omega<0.1$\ eV 
and the $\omega^{-4}$ extrapolation above 34 eV. We also made other types of extrapolation 
for $\hbar\omega < 0.1$ eV and check the difference of optical conductivity spectra, which 
will be discussed in the following sections.

\section{Resistivity measurements}
\begin{figure}
\includegraphics[width=8cm]{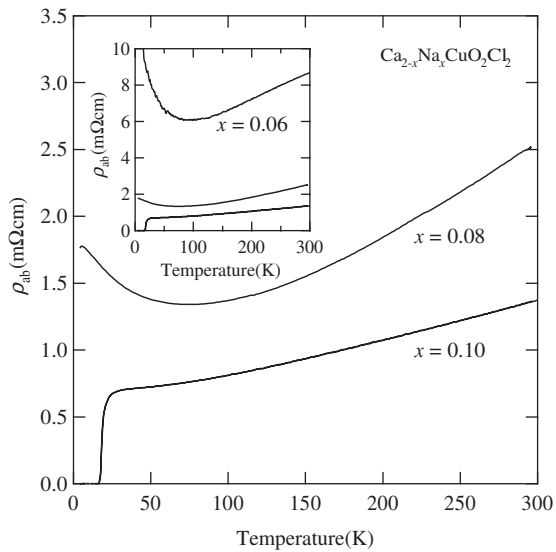}
\caption{\label{fig:Rab} The temperature dependence of the in-plane resistivity $\rho_{\rm 
ab}$ for CNCOC with $x=0.08$ and 0.10. The inset shows $\rho_{ab}$ of $x=0.06$, together 
with those of $x=0.08$ and 0.10.}
\end{figure}

\begin{figure}
\includegraphics[width=8cm]{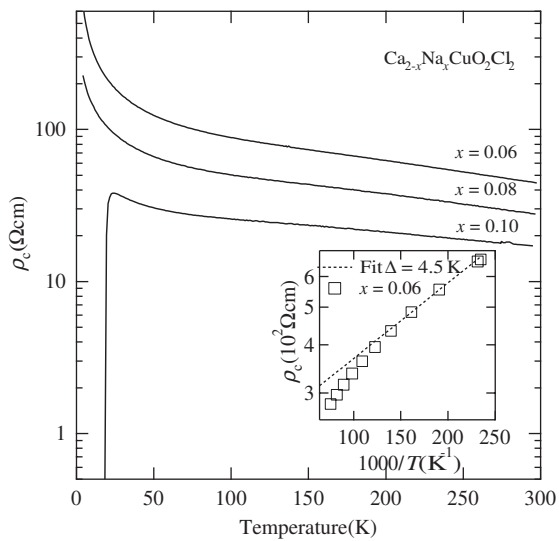}
\caption{\label{fig:Rc} The temperature dependence of the out-of-plane resistivity 
$\rho_{\rm c}$ for CNCOC. The inset is the Arrenius plot of $\rho_{\rm c}$ for $x=0.06$.}
\end{figure}

 Figure \ref{fig:Rab} shows the temperature dependence of the in-plane resistivity 
($\rho_{\rm ab}$) for CNCOC ($x \ge 0.06$). The absolute value and the temperature 
dependence of $\rho_{\rm ab}$ for CNCOC is similar to that of LSCO at the same doping 
level for $x=0.08$ and $x=0.10$.\cite{Komiya02} For $x=0.06$, however, the absolute value 
of $\rho_{\rm ab}$ is much larger than that of the LSCO counterpart. We speculate that the 
large value of $\rho_{\rm ab}$ for $x=0.06$ is caused by the mixing of the out-of-plane 
component, which often happens in the resistivity measurement of thin samples with large 
anisotropy. It should be noted that such a mixing barely affects the result of the out-of-plane 
resistivity. Figure \ref{fig:Rc} shows the temperature dependence of the out-of-plane resistivity 
($\rho_{\rm c}$) for CNCOC. The magnitude of $\rho_c$ at room temperature is about 50 
times larger than that of LSCO\cite{Komiya02} at the same doping level. As a result, the ratio 
of anisotropy in the resistivity ($\rho_{\rm c}/\rho_{\rm ab}$) amounts to $\sim 10^4$ in 
CNCOC. This larger absolute value of $\rho_c$ can be attributed to the smaller coupling of two 
adjacent CuO$_2$ planes in CNCOC, which are separated by the Ca(Na)Cl plane with a highly 
ionic character. By contrast, the temperature dependence of $\rho_c$ in CNCOC is smaller than 
that in LSCO. For example, at $x=0.10$, the resistivity ratio $\rho_{\rm c}$(50 K) / 
$\rho_{\rm c}$(290 K) is about 1.8 for CNCOC whereas 2.0 for LSCO.\cite{Komiya02} 
This discrepancy between the absolute value and the temperature dependence of $\rho_{\rm 
c}$ can hardly be explained by a conventional semiconductor model. One possible explanation 
is that the temperature dependence of $\rho_{\rm c}$ is dominated by the size of the so-called 
pseudogap as proposed previously.\cite{Takenaka94,Ong95} The pseudogap has been observed 
in various experiments, for example, NMR,\cite{Warren89} photoemission 
spectroscopy,\cite{White96} and even optical measurement.\cite{Basov96,Puchkov96} The 
size of the pseudogap should scale with the maximum $T_{\rm c}$ of each system, which is 28 
K for CNCOC and 38 K for LSCO in the present case. Therefore, it is expected that the size of 
the pseudogap in LSCO is larger than that in CNCOC, consistent with the temperature 
dependence of $\rho_{\rm c}$. As a more quantitative analysis, we estimated the size of the 
pseudogap from Arrenius plot of $\rho_{\rm c}(T)$. As shown in Fig. \ref{fig:Rc}, the 
activation energy of $\rho_{\rm c}(T)$, which scales with the size of the pseudogap, is about 
4.5 K in CNCOC. For BSSCO, the activation energy of $\rho_{\rm c}(T)$ is about 200 
K.\cite{Watanabe97} Such a huge difference of the size of the pseudogap estimated from 
$\rho_{\rm c}(T)$ may explain why $T_{\rm c}$ of CNCOC is so low compared with 
BSSCO.

\section{The doping dependence of the optical spectra}

\begin{figure}
\includegraphics[width=8cm]{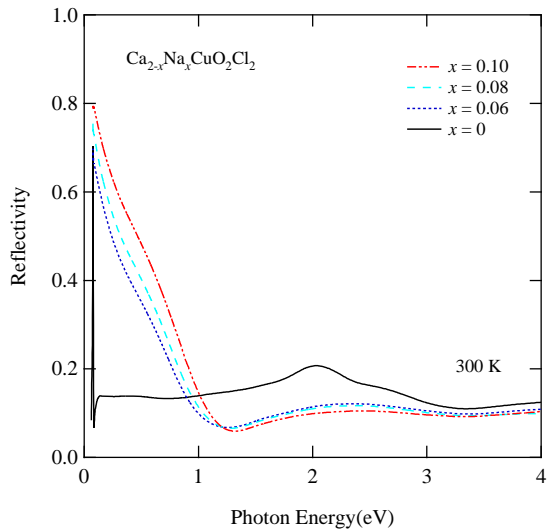}
\caption{\label{fig:Ref} (Color online) The doping dependence of reflectivity at room 
temperature.}
\end{figure}

\begin{figure}
\includegraphics[width=8cm]{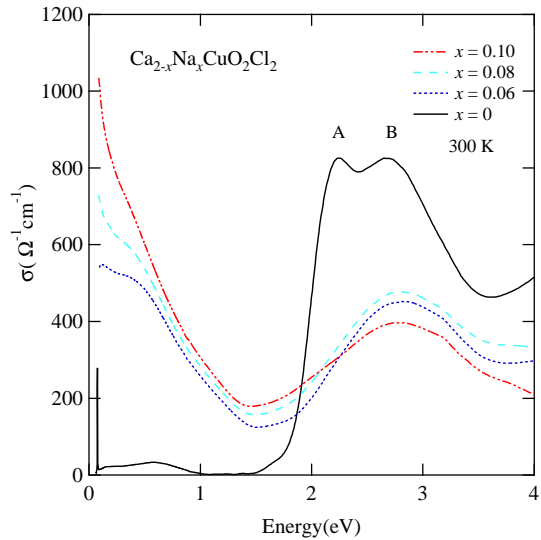}
\caption{\label{fig:Sig} (Color online) The doping dependence of optical conductivity at room 
temperature.}
\end{figure}

\begin{figure}
\includegraphics[width=8cm]{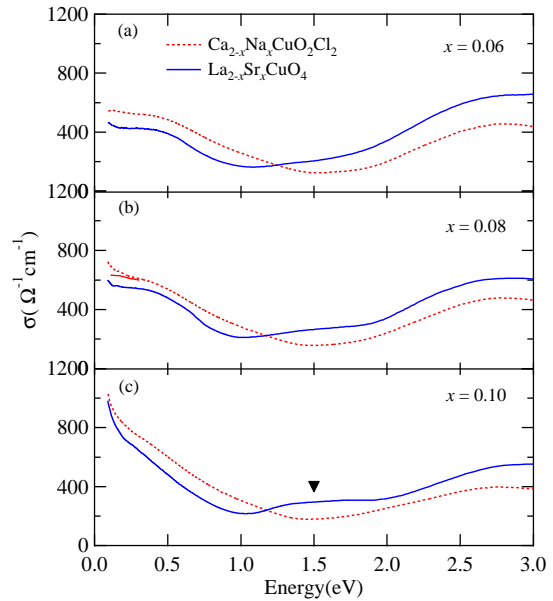}
\caption{\label{fig:Sig_LSCO} (Color online) Comparison of the optical conductivity between 
LSCO and CNCOC. The data of LSCO is from Ref. \onlinecite{Takenaka03}. The data of 
CNCOC (the dotted line), which is derived from the reflectivity spectrum with a Hagen-Rubens 
extrapolation, is previously shown in Fig. \ref{fig:Sig}. The optical conductivity spectrum 
derived from the reflectivity spectrum with a linear extrapolation below 0.1 eV is also plotted 
for $x$ = 0.08 (the solid line).}
\end{figure}

The doping dependence of the reflectivity of CNCOC is shown in Fig. \ref{fig:Ref}. The 
optical conductivity spectra derived from these reflectivity data are shown in Fig. \ref{fig:Sig}. 
The overall features of the spectrum and its doping dependence are the same as those of other 
cuprates: a sharp peak at 2.1 eV is suppressed and the quasi-Drude spectrum below 1.0 eV 
evolves with increasing $x$. However, there are several differences between CNCOC and 
LSCO. Figure \ref{fig:Sig_LSCO} compares the optical conductivity spectra of CNCOC and 
LSCO\cite{Takenaka03} with the same doping level. As can be seen, CNCOC always 
surpasses LSCO in the spectral weight of the Drude spectrum below 1.5 eV. Another difference 
is that there is a small peak at 1.5 eV (shown by a triangle) in the LSCO spectra, but such a peak 
is hardly seen in the CNCOC spectra.

In Fig. \ref{fig:Sig_LSCO}, the optical conductivity spectrum derived from the reflectivity 
spectrum with a linear extrapolation below 0.1 eV is also plotted for x = 0.08 (the solid line). As 
can be seen, there is a small difference between those with a Hagen-Rubens and a linear 
extrapolations below 0.2 eV. However, this difference is not large enough to qualitatively affect 
the following discussions.

To make more quantitative discussions about the difference and the similarity of the spectra 
between CNCOC and LSCO, the effective number of electrons, $N_{\rm eff}$ was estimated 
in the following way, 

\begin{equation}
N_{\rm eff}=\frac{2mV}{{\pi} e^2}\int_0^{\omega_c}  {\sigma}({\omega}^{'}) 
d{\omega}^{'}  \label{eq:Neff}
\end{equation}

\begin{figure}
\includegraphics[width=8cm]{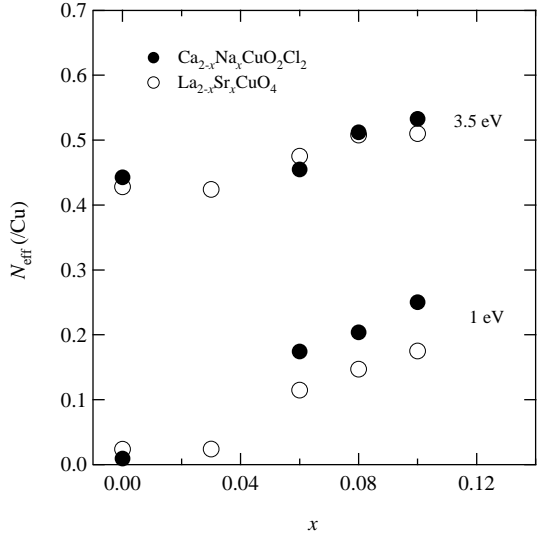}
\caption{\label{fig:Neff} $N_{\rm eff}$ of CNCOC and LSCO with the cut-off energy $\hbar 
\omega _{c}=1$ eV and 3.5 eV.}
\end{figure}

Figure \ref{fig:Neff} plots $N_{\rm eff}$ with the cut-off energy $\hbar \omega _{c}=1$ eV 
and 3.5 eV. It is noticeable that $N_{\rm eff}$ with $\hbar \omega _{c}=3.5$ eV is almost the 
same between CNCOC and LSCO for the same doping level. This indicates that the spectral 
weight below 3.5 eV is governed by a common component of CNCOC and LSCO, i.e., the 
CuO$_{2}$ plane, and La, Sr, Ca, and Na do not largely contribute to the spectrum below 3.5 
eV. Thus, it can be concluded that the 1.5 eV peak existing only in the LSCO spectra also comes 
from the excitation in the CuO$_{2}$ plane. 

Both CNCOC and LSCO have the same structure of the single CuO$_{2}$ plane, and the only 
difference in the CuO$_{2}$ plane between these two systems is the buckling of the 
CuO$_{6}$ octahedra and a stripe instability, both of which exist only in LSCO. It should be 
noted here that the buckling of the CuO$_{6}$ octahedra disappears for $x>0.20$ in 
LSCO,\cite{Fleming87} where the 1.5 eV peak still survives.\cite{Uchida91} Therefore, it is 
plausible to assign the 1.5 eV peak in the optical conductivity of LSCO as the excitation 
associated with the charge stripe. This assignment can also explain why the 1.5 eV peak does 
not exist in either YBCO\cite{Cooper93} or BSCCO.\cite{Terasaki90} 

\begin{figure}
\includegraphics[width=8cm]{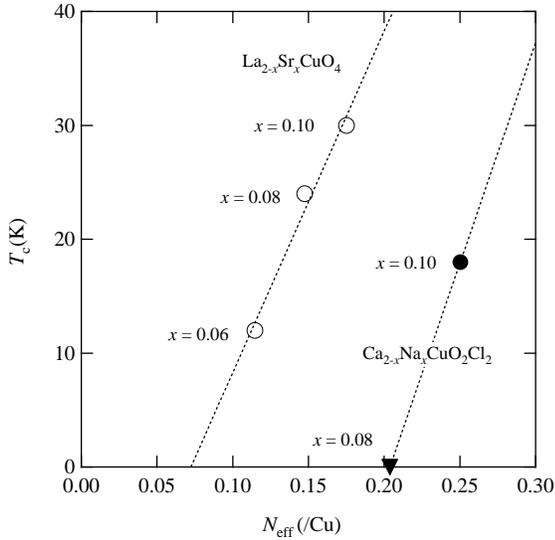}
\caption{\label{fig:Tc}The relation between $N_{\rm eff}$ at 1 eV and the superconducting 
transition temperature $T_{\rm c}$ of LSCO and CNCOC. Dotted lines are guide to eyes.}
\end{figure}

In Fig. \ref{fig:Neff}, it is also found that $N_{\rm eff}$ with $\hbar \omega _{c}=1$ eV, 
which corresponds to the Drude weight of the systems, of CNCOC is larger than that of LSCO. 
This result is counterintuitive, if one recalls the phase diagram of these two systems; LSCO 
becomes superconducting for the smaller value of $x$ ($\ge 0.06$) than CNCOC ($\ge 
0.09$).\cite{Hiroi96} This is more clearly seen in Fig. \ref{fig:Tc}, where $N_{\rm eff}$ and 
the superconducting transition temperature $T_{\rm c}$ of each sample are plotted. As can be 
seen, CNCOC and LSCO follow the different trend, indicating that $N_{\rm eff}$ is by no 
means the dominant parameter of $T_{\rm c}$.

 Since these two systems, CNCOC and LSCO, have a similar crystal structure (the single 
CuO$_{2}$ plane), the result is rather surprising. One possible explanation is that not all of the 
Drude spectrum below 1.0 eV contributes to the superconductivity. In other words, the spectral 
weight that condensates to the superfluid, which dominates the transition temperature, is only a 
fraction of the spectral weight below 1.0 eV. This interpretation suggests the two-fluid nature of 
the quasi-Drude spectrum below 1.0 eV, but how the spectrum is divided into two components 
remains unclear.
 
\section{The temperature dependence of the optical spectra}

\begin{figure}
\includegraphics[width=8cm]{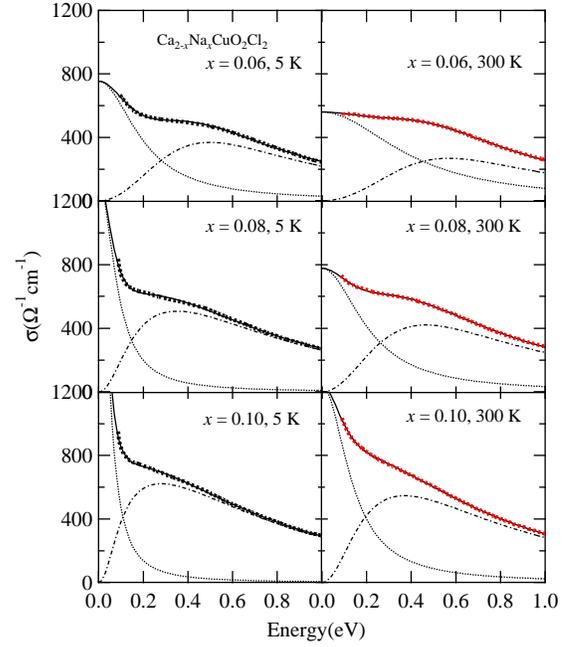}
\caption{\label{fig:2ComFit} (Color online) The closed symbols are the optical conductivity 
obtained from the reflectivity spectra. The solid lines are the result of the fitting by the sum of a 
Drude and a Lorentzian components. The dotted lines are the Drude components, and the 
dot-dashed lines are the Lorentzian components.}
\end{figure}

Figure \ref{fig:2ComFit} shows the temperature dependence of the optical conductivity spectra 
for CNCOC with $x=0.06$, 0.08, and 0.10 (solid symbols). It is well known that the Drude-like 
spectrum below 1 eV in the cuprate superconductors cannot be fitted by a single Drude form, 
and there have been a lot of arguments about whether the one-component model (the so-called 
extended Drude model) or the two-component model (the Drude and Lorentzian model) is 
appropriate to explain such a spectrum. Here, we analyze the experimental data of CNCOC in 
both ways. 

First, the spectra were analyzed by the two-component model,\cite{Thomas88} i.e., the sum of 
a Drude component and a Lorentzian in the following way;

\begin{equation}
\sigma(\omega)=\sigma_{\rm D}(\omega) + \sigma_{\rm L}(\omega)  
\label{eq:Fit2Com}
\end{equation}

\begin{equation}
\sigma_{\rm D}(\omega) = \frac{\omega_{\rm D}^2}{4\pi}\frac{\Gamma_{\rm 
D} }{\omega^2+\Gamma _{\rm D}^2}  \label{eq:FitDrude}
\end{equation}

\begin{equation}
\sigma_{\rm L}(\omega) =\frac{S_{\rm L}\omega_{\rm L}^2 }{4\pi}\frac{\omega 
^2\Gamma _{\rm L}}{(\omega^2-\omega_{\rm L}^2)^2+\omega^2\Gamma_{\rm L}^2} 
\label{eq:FitLorentz}
\end{equation}

\begin{figure}
\includegraphics[width=8cm]{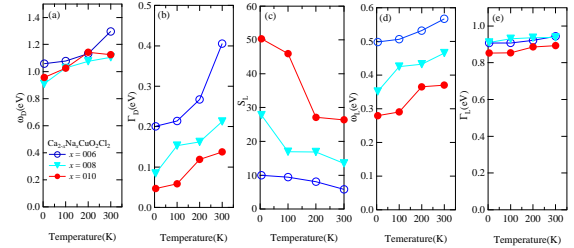}
\caption{\label{fig:2ComPara} (Color online) The doping- and the temperature-dependence of 
the five parameters, $\omega_{\rm D}$, $\Gamma_{\rm D}$, $S_{\rm L}$, $\omega_{\rm 
L}$, and $\Gamma_{\rm L}$ derived from the fitting of the optical conductivity spectrum. 
The fitting function is shown in the text [Eq. (\ref{eq:Fit2Com}), (\ref{eq:FitDrude}), 
(\ref{eq:FitLorentz})]. }
\end{figure}

Here, there are five fitting parameters (two for the Drude and three for the Lorentzian), the 
plasma frequency of the Drude component $\omega_{\rm D}$, the scattering rate of the Drude 
component $\Gamma_{\rm D}$, the oscillator strength of the Lorentzian $S_{\rm L}$, the 
peak position of the Lorentzian $\omega _{\rm L}$, and the scattering rate of the Lorentzian 
$\Gamma_{\rm L}$. The result of the fitting to each spectrum is quite satisfactory, as shown 
by the solid lines in Fig. \ref{fig:2ComFit}. The doping- and the temperature-dependence of the 
five parameters are summarized in Fig. \ref{fig:2ComPara}. As can be seen (1) $\omega_{\rm 
D}$ barely changes with doping, but decreases with decreasing temperature (2) 
$\Gamma_{\rm D}$ decreases with increasing doping and with decreasing temperature (3) 
$S_{\rm L}$ increases with increasing doping but does not change with temperature (4) 
$\omega_{\rm L}$ decreases with increasing doping and with decreasing temperature (5) 
$\Gamma_{\rm L}$ barely changes with doping and temperature. 

There are various theoretical studies on the strongly correlated systems with 
doping.\cite{Tohyama91,Pruschke93,Dagotto92} However, the present experimental results 
have several significant discrepancies with those theoretical studies. First, most of the theories 
predict the increase of the Drude frequency $\omega_{\rm D}$ (or Drude weight) with 
increasing doping and decreasing temperature, both of which are inconsistent with the present 
experimental result [Fig. \ref{fig:2ComPara}(a)]. Second, the localized state, which is 
represented by a Lorentzian form, usually shifts to a higher energy with decreasing temperature. 
This behavior is also true for the theoretical studies of the strongly correlated system with 
infinite dimensions.\cite{Pruschke93} However, the present experimental result indicates that 
$\omega _{\rm L}$ rather decreases with decreasing temperature [Fig. 
\ref{fig:2ComPara}(d)], hard to reconcile with the theories. The similarity between the doping- 
and the temperature-dependence of $\Gamma_{\rm D}$ [Fig. \ref{fig:2ComPara}(b)] and 
$\omega _{\rm L}$ [Fig. \ref{fig:2ComPara}(d)] suggests that the spectrum assigned to a 
Lorenztian component (a localized state) in this fitting is not really a localized state, but a part 
of a Drude component (an itinerant state).  In other words, the present analysis suggests that 
the one-component model is more plausible to fit the data of CNCOC than the two-component 
model. 

\begin{figure}
\includegraphics[width=8cm]{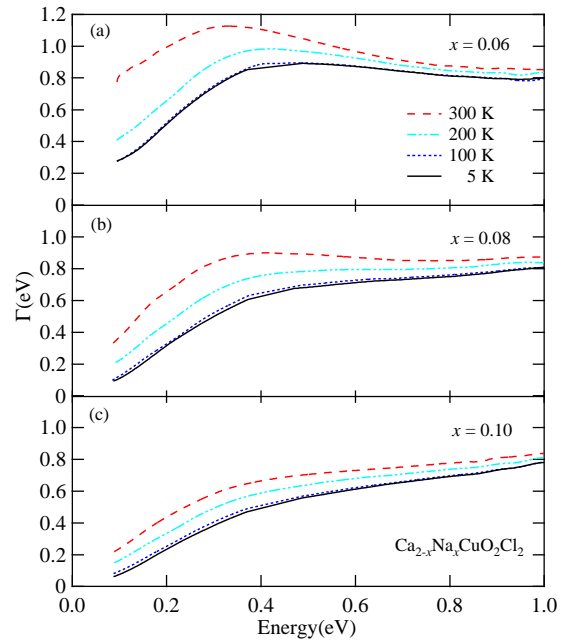}
\caption{\label{fig:Tau1} (Color online) The $\omega$-dependence of the scattering rate 
$\Gamma$ derived from the extended Drude analysis.}
\end{figure}

We also analyze the data by the one-component model, i.e., the extended Drude 
model.\cite{Rotter91} In this model, all of the spectrum below 1.0 eV is assigned to an 
itinerant state, but the effective mass and the scattering rate are both $\omega$-dependent and 
are derived by the following expression,

\begin{equation}
\Tilde{\sigma}(\omega)=\frac{4{\pi}ne^2}{m^{*} 
(\omega)}\frac{i}{{\omega}+i\Gamma(\omega) }, \label{eq:ExtendedDrude}
\end{equation}
here $\Tilde{\sigma}(\omega)$ is the complex optical conductivity.
The result of the $\Gamma(\omega)$ at various temperatures at each $x$ is shown in Fig. 
\ref{fig:Tau1}. There is a strong $\omega$ dependence of $\Gamma(\omega)$ in each 
$x$ and temperature. Particularly at low frequency, the scattering rate has a term almost 
proportional to $\omega$, in such a way that $\Gamma(\omega) = \Gamma_{0}+C\omega$, 
which is a common behavior of cuprate superconductors.\cite{Rotter91} As can be seen in Fig. 
\ref{fig:Tau1}, the $\omega$-coefficient in $\Gamma(\omega)$ at low frequency, $C$, is 
almost temperature independent, but only the constant term $\Gamma_{0}$ decreases with 
decreasing temperature. It can also be seen in Fig. \ref{fig:Tau1} that $\Gamma(\omega)$ is 
saturated for $\hbar\omega \gtrsim 0.4$ eV. We speculate that this behavior is similar to the 
behavior of resistivity saturation observed in the dc resistivity at high 
temperatures.\cite{Takagi92} This issue is discussed in the next section.

\begin{figure}
\includegraphics[width=8cm]{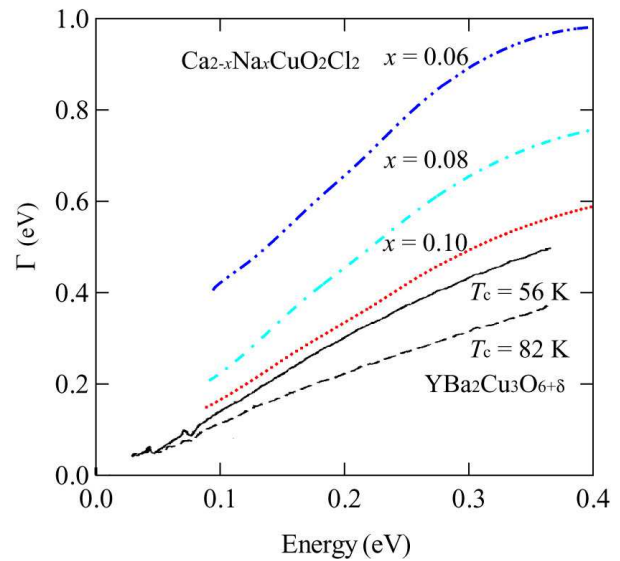}
\caption{\label{fig:Tau} (Color online) Comparison of $\Gamma(\omega)$ between CNCOC 
and YBCO. The data of CNCOC is derived from the optical conductivities at 200 K. The data of 
YBCO is from Ref. \onlinecite{Rotter91}. }
\end{figure}

Figure \ref{fig:Tau} compares the $\omega$ dependence of the scattering rate, 
$\Gamma(\omega)$, of CNCOC and YBCO.\cite{Rotter91} As can be seen, 
$\Gamma(\omega)$ of CNCOC with $x=0.10$ and YBCO with $T_{\rm c}=56$ K is almost 
the same. It should be noted that the ratio of $T_{\rm c}$ for CNCOC with $x=0.10$ (18 K) to 
the maximum $T_{\rm c}$ of the same series (28 K for $x=0.15$) , which is a good measure of 
the hole concentration, is $\sim$ 0.6, and this value is almost the same as that of YBCO with 
$T_{\rm c}=56$ K. This indicates that $\Gamma(\omega)$ is the same for the samples with 
the same hole concentration, even though the systems are different. Such a universality of 
$\Gamma(\omega)$ with the change of the hole concentration should be an intrinsic nature of a 
CuO$_2$ plane with a tetragonal lattice.

\section{Discussion}

As shown in Fig. \ref{fig:Sig}, there are distinct two peaks (A and B) in the optical 
conductivity spectrum of the parent compound between 2 and 3 eV, both of which can be 
assigned to the CT excitation. With hole doping, however, only the A peak disappears but the B 
peak survives. Such a doping dependence of the double-peak structure of the CT excitation has 
not been discussed so far, mainly because the double-peak structure in the parent compound 
itself is not so clear in other cuprate superconductors. 

There are several explanations for the double peaks of the CT excitation. One explanation is to 
ascribe them to the two kinds of holes, those in the Zhang-Rice singlet band (ZRB) and in the 
non-bonding oxygen $2p$ band (NBB).\cite{Choi99} Here, it should be emphasized that both 
two hole bands can be observed in the ARPES spectrum of the parent compound, and that both 
bands survive even in the hole-doped samples.\cite{Ronning03} Therefore, only the splitting of 
ZRB and NBB cannot explain the fact that the A peak disappears with hole doping in the optical 
spectrum. We speculate that the excitonic effect between oxygen $2p$ holes and Cu 
$3d$ electrons is essential to the doping dependence of the optical spectrum, which inherently 
does not exist in the photoemission process. 

 Next, the $\omega$-dependence of the scattering rate, $\Gamma(\omega)$, is saturated at 
large $\omega$ as shown in Fig. \ref{fig:Tau1}. It is noticeable that the saturation value of 
$\Gamma$ is almost the same for any temperature and $x$, i.e., $\Gamma \sim 0.8$ eV. This 
behavior is similar with the so-called resistivity saturation, which occurs in the dc resistivity at 
high temperatures under the condition $k_{\rm F}\ell \simeq 1$, where $k_{\rm F}$ is the 
Fermi wavenumber and $\ell$ is the mean free path of the carrier. Considering the fact that the 
scattering rate $\Gamma$ has both the temperature dependence and the $\omega$ dependence, 
$\Gamma(\omega)$ should show a saturation behavior just as $\Gamma(T)$ does in the dc 
resistivity. Here we estimate the saturation value of $\Gamma(\omega)$ as follows. Using the 
relations ${\ell}=v_{\rm F}\tau$ where $v_{\rm F}$ is Fermi velocity and $\tau$ is the 
relaxation time and $v_{\rm F}={\hbar}k_{\rm F}/m^*$, we can rewrite the relation $k_{\rm 
F}\ell =1$ as $\Gamma=\hbar^2k_{\rm F}/m^*$. In the two-dimensional system, $k_{\rm 
F}$ is given by $(2{\pi}nd)^{1/2}$($d$ is the inter-plane distance and $n$ is the carrier 
density), and thus $\Gamma=n/m^* \times 2\pi\hbar^2d$. The unknown parameter, $n/m^*$, 
was estimated from the experimentally obtained effective number of electrons at 1eV. The 
$\Gamma$ value thus derived becomes 0.56 eV, 0.66 eV and 0.80 eV for $x=0.06$, 0.08 and 
0.10, respectively. These are in good agreement with the experimentally estimated values of 
$\Gamma$ at the saturation point, indicating that the saturation of the $\Gamma(\omega)$ has 
the same origin as that of $\Gamma(T)$. 

We also discuss another possible interpretation of the structure in $\Gamma(\omega)$: 
In the ARPES data of the same compounds, the spectral weight below 0.4 eV is heavily suppressed, 
particularly around $(\pi,0)$ point in the $k$ space. This behavior suggests either 
(a) a pseudogap ($\Delta = 0.4$ eV) opens on the large Fermi surface around $(\pi, 0)$ point, or 
(b) only small hole pockets exist around $(\pi/2,\pi/2)$ point. If (a) the pseudogap picture is correct, 
there should be an excitation between the pseudogap in the optical spectrum, 
which could appear around $\Delta - 2\Delta$, and the ``shoulder'' around 0.4 eV 
in $\Gamma(\omega)$ can be attributed to the excitation. However, this scenario is rather unlikely, 
because (1) as shown in Fig. \ref{fig:Tau}, the $\Gamma(\omega)$ below 0.4 eV for $x=0.10$ coincides with 
that of YBCO with $T_{\rm c} = 56 K$, which does not have such a large pseudogap, 
and (2) the energy scale of the structure in $\Gamma(\omega)$ barely changes, or rather increases, 
with increasing hole concentration, inconsistent with the behavior of the conventional pseudogap 
whose energy decreases with hole doping. On the other hand, on the basis of (b) the hole-pocket picture, 
the possible final state of the optical spectrum is the upper Hubbard band, which is $\sim 2$ eV 
above the hole band, and thus, there would be no excitation below 1eV except for a Drude response 
in the optical spectrum. This is consistent with the interpretation of the optical conductivity spectra 
with the $\Gamma(\omega)$ saturation, as discussed above. This picture implies that the hole doping into 
Ca$_{2}$cCuO$_{2}$Cl$_{2}$ can be described as a rigid band shift without any large reconstruction 
of the valence and the conduction band in the underdoped region.

\section{Summary}
In this paper, we report the resistivity and reflectivity measurement of 
Ca$_{2-x}$Na$_x$CuO$_2$Cl$_2$ (CNCOC), which has a purely tetragonal CuO$_2$ plane 
and thus, is the best system to investigate the charge dynamics of tetragonal lattice with strong 
electron correlation. It was found that the absolute value of the out-of-plane resistivity of 
CNCOC is much larger than that of LSCO owing to the ionic character of the (Ca,Na)Cl plane, 
though its temperature dependence is smaller for CNCOC. This discrepancy suggests that the 
temperature dependence of the out-of-plane resistivity is dominated by the opening of a 
pseudogap. It was also found that the doping dependence of the in-plane optical conductivity 
spectra of CNCOC is similar with those of other cuprate superconductors, but a careful 
comparison of the spectra between CNCOC and La$_{2-x}$Sr$_x$CuO$_4$ (LSCO) clarifies 
that (1) there is a small peak around 1.5 eV between the charge-transfer peak(2 eV) and a 
quasi-Drude peak (below 1 eV) only in LSCO (2) the Drude weight below 1 eV of CNCOC is 
always larger than that of LSCO at the same doping level, though the superconducting transition 
temperature is lower for CNCOC. The 1.5 eV peak existing only in LSCO can be attributed to 
the charge stripe in LSCO. The larger Drude weight  and lower transition temperature implies 
that only a part of the Drude weight below 1 eV contributes to the superconductivity.

The temperature dependence of the optical conductivity spectra of CNCOC has been analyzed 
both by the two-component model (Drude+Lorentzian) and by the one-component model 
(extended Drude analysis). Five fitting parameters can be obtained by the two-component model, 
but the doping and temperature dependence of those parameters are not in good agreement with 
the theoretical predictions. On the other hand, it was found that the $\omega$-dependence of 
$\Gamma$ derived from the extended Drude analysis shows a universal change with doping for 
different systems. It was also found that $\Gamma(\omega)$ shows a saturation behavior 
above 0.4 eV, which has the same origin of the resistivity saturation at high temperatures, i.e., 
$k_{\rm F}\ell$ cannot be smaller than unity. Finally, the absence of the structure below 1.0 
eV in the present optical conductivity spectra, together with the result of the ARPES 
measurement that the spectral weight around $(\pi,0)$ point is suppressed below 0.4 eV, 
suggests that only small pockets exist around $(\pi/2,\pi/2)$ point in the underdoped regime of 
CNCOC.

\begin{acknowledgments}
The authors would like to thank T. Hanaguri for technical assistance on resistivity 
measurements and helpful discussions, K. Takenaka for sending us his data of LSCO, and S. 
Miyasaka and Y. Tokura for their help with the measurement at UV-SOR.
This work was partly supported by a Grant-in-Aid for The 21st Century COE Program (Physics 
of Self-organization Systems) at Waseda University from the Ministry of Education, Sports, 
Culture, Science and Technology of Japan
\end{acknowledgments}

\bibliography{CNCOC}

\end{document}